\documentclass[fleqn,twoside]{article}
\usepackage[headings]{espcrc2}

\readRCS
$Id: espcrc2.tex,v 1.2 2004/02/24 11:22:11 spepping Exp $
\usepackage{graphicx, balance}
\usepackage[figuresright]{rotating}

\newcommand{\AmS}{{\protect\the\textfont2
  A\kern-.1667em\lower.5ex\hbox{M}\kern-.125emS}}

\title{\textbf{Text Classification and Distributional features techniques in Datamining and Warehousing}}

\author{Srikanth Bethu\address[DCSE]{Assistant Professor,Department of Computer Science and Engineering, Holymary Institute of Technology and Science, JNTU Hyderabad - 501 301 India,
Contact: srikanthbethu@gmail.com \\},
G Charless Babu\address{Professor, Department of Computer Science and Engineering, Holymary Institute of Technology and Science},J Vinoda\address{Department of Computer Science and Engineering, Holymary Institute of Technology and Science, JNTU Hyderabad},E Priyadarshini\address{Assistant Professor,Department of Computer Science and Engineering, Holymary Institute of Technology and Science, JNTU Hyderabad}M Raghavendra rao\address{Assistant Professor,Department of Computer Science and Engineering, Holymary Institute of Technology and Science, JNTU Hyderabad.}}

\runtitle{Preparation of International Journal of Information Processing Paper in Two-Column Format}
\runauthor{Srikanth Bethu, et al.,}

\begin{document}
\begin{abstract}

 Text Categorization is traditionally done by using the term frequency and inverse document frequency.This type of method is not very good because, some words which are not so important may appear in the document .The term frequency of unimportant  words may increase and document may be classified in the wrong category.For reducing the error of classifying of documents in wrong category. The Distributional features are introduced. In the Distribuional Features, the Distribution of the words in the whole document is analyzed. Whole Document is very closely analyzed for different measures like FirstAppearence, Last Appearance, Centriod, Count, etc.The measures are calculated and they are used in tf*idf equation and result is used in k- nearest neighbor and K-means  algorithm for classifying the documents.
.  
\end{abstract}

\maketitle

\section{INTRODUCTION}
Text classification is the task of automatically classifying set of documents into categories from a predefined set. This task has several applications selective distribution of information to information consumers, spam filtering, and identification of document type. Automated text classification is good because it frees organizations from the need of manually organizing document. Text classification has gained importance because of the increased amount of documents over the years. Text documents should be categorized according to its contents.
The aim of the paper is to classify the document in the correct category. That means the document should be classified into correct type of document to which it belongs, for example the computer science type of document should be classified as computer science type of document.

\subsection{Problem definition}
\label{1}
The classification of the text is traditionally done by the term frequency and the inverse document frequency, this method has lot of problems. Term frequency cannot classify the documents in the correct category because the unimportant words may appear more number of times and the document may be classified to wrong category. The below example will explain the problem.

\subsection{Example:Bill Gates philanthropy costs him richest-man title}
\label{2}
\renewcommand{\labelenumi}{(\roman{enumi})}
\begin{enumerate}
\item Bill Gates attends a session at the World Economic Forum (WEF) in Davos January 28, 2011.

\item Bill Gates didn't lose his title as the world's richest man last year; he gave it away by plowing billions into his charitable foundation, experts say.

\item Forbes will release its 2011 billionaires list on Wednesday and Gates, investor Warren Buffett and last year's richest man, Mexican tycoon Carlos Slim, will almost certainly be in the top three. The trio has topped the list for the past five years.

\item Lincoln said Gates was currently worth about \$49 billion, behind Slim, whose fortune he estimated at \$60 billion. Buffett, also a philanthropist, is now worth some \$47 billion.

\item But had Gates not given away any money, he would be worth \$88 billion, Lincoln said.

\item Gates and his wife Melinda have so far given \$28 billion to their foundation, the largest in the United States.

\item Forbe's 2010 billionaires list put Gates' fortune at \$53 billion, but he was knocked into second spot by Slim's \$53.5 billion, losing the crown for only the second time since 1995.

\end{enumerate}
In the above example the word gates appeared 4 times and the term frequency is more since the document will be classified as gates but the document is about Bill Gates but not about Gate, because the gate means an entrance.

\subsection{Distributional Features}
\label{3}
The distributional features [9] are the emerging technique for the classification of the document. The distributional features will take into account for the complete contents of the document to be classified. The distributional features will study the distribution of the words in the document and classify according to that.
The distributional features will take into account of the following main points.
\renewcommand{\labelenumi}{(\roman{enumi})}
\begin{enumerate}

\item The words occurring in the title and in the introduction of the document are very important.

\item The words occurring in the conclusion and in the last part of the document are very important.

\item The words occurring in the middle part of document are not important.

\end{enumerate}
The new measure or the technique discovered is the compactness of the appearances of a word. The compactness [9] measures whether the appearances of a word concentrate in a specific part of a document or spread over the whole document. If the word is in specific part of document, the word is considered as compact. If the word is spread over the document, the word is considered as less compact. This consideration is motivated by the following facts. A document usually contains several parts. If the appearances of a word are less compact, the word is more likely to appear in different parts and more likely to be related to the theme of the document.
The compactness of the appearances of a word shows that the less compactly a word appears, the more important the word is and the position of the first appearance of a word shows that the earlier a word is mentioned, the more important this word is.

\section{Designing and Implementation of Text Classification Techniques}

\subsection{Distributional Features}
\label{1}
The distributional features means the word distribution in the document and their appearance in the document entire document is taken into account. If an important word is occurred only few times then also the distributional features [9] will correct and good classification.
The distributional features will make into account various measures like first appearance, last appearance, count, centroid, compactness. They will calculate the importance of the word in the document and the measures will classify the document correctly.

\subsection{Text Classification}
\label{2}
The task is to assign an electronic document to one or more categories, based on its contents. Document classification tasks can be divided into two sorts: supervised document classification [4] where some external mechanism (such as human feedback) provides information on the correct classification for documents, and unsupervised document classification (also known as document clustering), where the classification must be done entirely without reference to external information. There is also a semi-supervised document classification, where parts of the documents are labeled by the external mechanism.

\subsection{k-Nearest Neighbor Algorithm}
\label{3}
In pattern recognition, the k-nearest neighbors algorithm (k-NN) is a method for classifying objects based on closest training examples in the feature space. k-NN is a type of instance-based learning, or lazy learning where the function is only approximated locally and all computation is deferred until classification. The k-nearest neighbor algorithm is amongst the simplest of all machine learning algorithms: an object is classified by a majority vote of its neighbors, with the object being assigned to the class most common amongst its k nearest neighbors (k is a positive integer, typically small). If k = 1, then the object is simply assigned to the class of its nearest neighbor.

The same method can be used for regression, by simply assigning the property value for the object to be the average of the values of its k nearest neighbors. It can be useful to weight the contributions of the neighbors, so that the nearer neighbors contribute more to the average than the more distant ones.

The neighbors are taken from a set of objects for which the correct classification or, in the case of regression, the value of the property is known. This can be thought of as the training set for the algorithm, though no explicit training step is required. The k-nearest neighbor algorithm is sensitive to the local structure of the data.
Nearest neighbor rules in effect compute the decision boundary in an implicit manner. It is also possible to compute the decision boundary itself explicitly, and to do so in an efficient manner so that the computational complexity is a function of the boundary complexity.

Algorithm:
The training examples are vectors in a multidimensional feature space, each with a class label. The training phase of the algorithm consists only of storing the feature vectors and class labels of the training samples.
In the classification phase, k is a user-defined constant, and an unlabelled vector (a query or test point) is classified by assigning the label which is most frequent among the k training samples nearest to that query point.

Parameter selection:
The best choice of k depends upon the data; generally, larger values of k reduce the effect of noise on the classification, but make boundaries between classes less distinct. A good k can be selected by various heuristic techniques, for example, cross-validation. The special case where the class is predicted to be the class of the closest training sample (i.e. when k = 1) is called the nearest neighbor algorithm.

The nearest neighbor algorithm has some strong consistency results. As the amount of data approaches infinity, the algorithm is guaranteed to yield an error rate no worse than twice the Bayes error rate (the minimum achievable error rate given the distribution of the data) . k-nearest neighbor is guaranteed to approach the Bayes error rate, for some value of k (where k increases as a function of the number of data points). Various improvements to k-nearest neighbor methods are possible by using proximity graphs.
Example of the KNN Algorithm.

\begin{figure}
\centering
\resizebox{8cm}{6cm}{\includegraphics{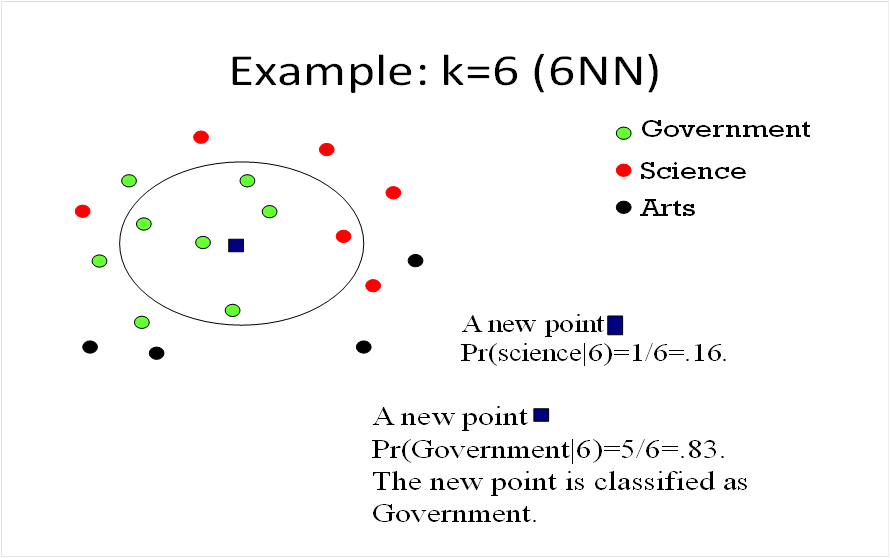}}
\caption{K-Nearest Neighbor Algorithm Example}
\end{figure}

 An improved KNN text classification algorithm[1], based on clustering center is proposed .The  training sets are compressed and the samples near by the border are deleted.The training sample sets of each category are clustered by k-means clustering algorithm, and all cluster centers are taken as the new training samples. A weight value is introduced, which indicates the importance of each training sample according to the number of samples in the cluster that contains this cluster center. Finally, the modified samples are used to accomplish KNN text classification.

\subsection{K means Algorithm:}
\label{4}
\renewcommand{\labelenumi}{(\roman{enumi})}
\begin{enumerate}
\item k-means clustering is a method of cluster analysis which aims to partition n observations into k clusters in which each observation belongs to the cluster with the nearest mean.

\item Given a set of observations (x1, x2, …, xn), where each observation is a d-dimensional real vector, then k-means clustering aims to partition the n observations into k sets (k < n) ,S={S1, S2, …, Sk}.

\item The procedure follows a simple and easy way to classify a given data set through a certain number of. The main idea is to define k centroids, one for each cluster. These centroids shoud be placed in a careful way because of different location causes different result. So, the better choice is to place them as much as possible far away from each other. The next step is to take each point belonging to a given data set and associate it to the nearest centroid. When no point is pending, the first step is completed and an early groupage is done. At this point we need to re-calculate k new centroids of the clusters resulting from the previous step. After we have these k new centroids, a new binding has to be done between the same data set points and the nearest new centroid. A loop has been generated. As a result of this loop we may notice that the k centroids change their location step by step until no more changes are done. In other words centroids do not move any more.
\end{enumerate}
Finally, this algorithm aims at minimizing an objective function, in this case a squared error function. The objective function.

\begin{equation}\label{1}
\displaystyle J=\sum_{j\mathop=1}^k\sum_{j\mathop=1}^x\amalg x_i^(j)-c_j\amalg^2.
\end{equation}

where ||xi(j) –cj||2 is a chosen distance measure between a data point xi(j) and the cluster centre cj, is an indicator of the distance of the n data points from their respective cluster centres.

The algorithm is composed of the following steps:
\renewcommand{\labelenumi}{(\roman{enumi})}
\begin{enumerate}
\item Randomly choose k data points to be the initial centroids, cluster centers.
\item Assign each data point to the closest centroid.
\item Re-compute the centroids using the current cluster memberships.
\item If a convergence criterion is not met, go to 2).
\end{enumerate}

\begin{enumerate}
\item Example:
\end{enumerate}

\begin{figure}
\centering
\resizebox{8cm}{6cm}{\includegraphics{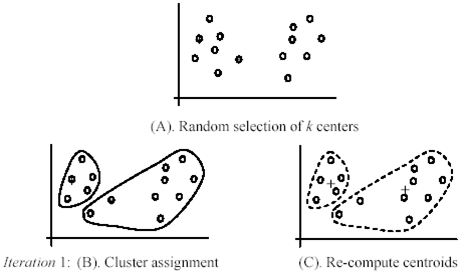}}
\caption{K means algorithm Example}
\end{figure}

\begin{figure}
\centering
\resizebox{8cm}{6cm}{\includegraphics{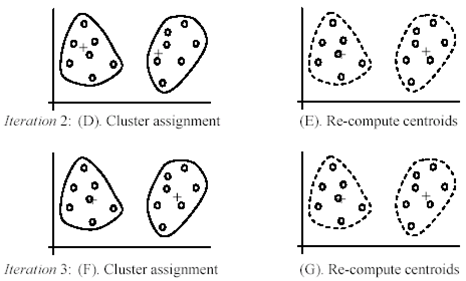}}
\caption{K means algorithm Example}
\end{figure}

The k-means clustering technique has been implemented which is like with hierarchical initial set (HKM). The goal is to prove that clustering document sets do enhancement precision on information retrieval systems, since it was proved by Bellot and El-Beze on French language.

\section{Experimental Results }

The corpus consists of 30 documents, which need to be classified. The 15 documents are of institute for electrical and electronic engineers, related to computer science engineering. The remaining 15 documents are of conference papers of various topics like medical, civil engineering, etc.

The thirty documents are taken as test documents and, the performance is calculated using both traditional method using term frequency and distributional feature that is using compactness, using the precision and recall measures. The precision of the traditional method and compactness give the same result, but the recall measure of distributional features gives the better result than traditional method of classifying the documents.

Distributional features give more accurate classification than the traditional term frequency method of classification of documents. Therefore distributional features are useful for classification of the documents.

\begin{figure}
\centering
\resizebox{8cm}{6cm}{\includegraphics{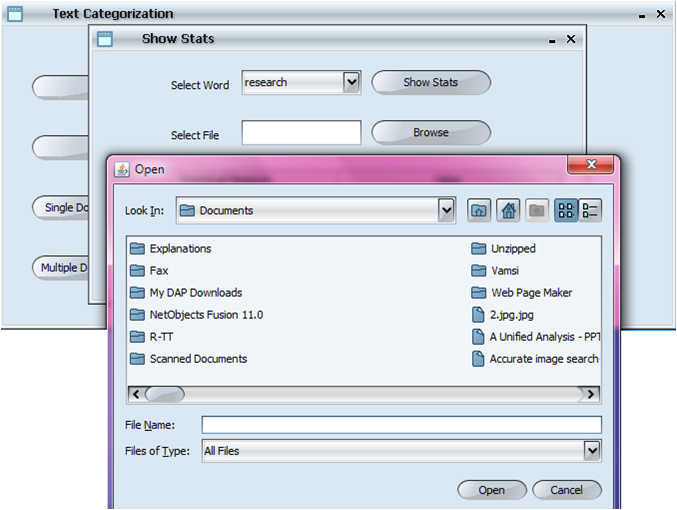}}
\caption{Selection of document from the folder}
\end{figure}

Figure shows the selection of document from the folder, the document which is required for analysis can be uploaded for system.

\begin{figure}
\centering
\resizebox{8cm}{6cm}{\includegraphics{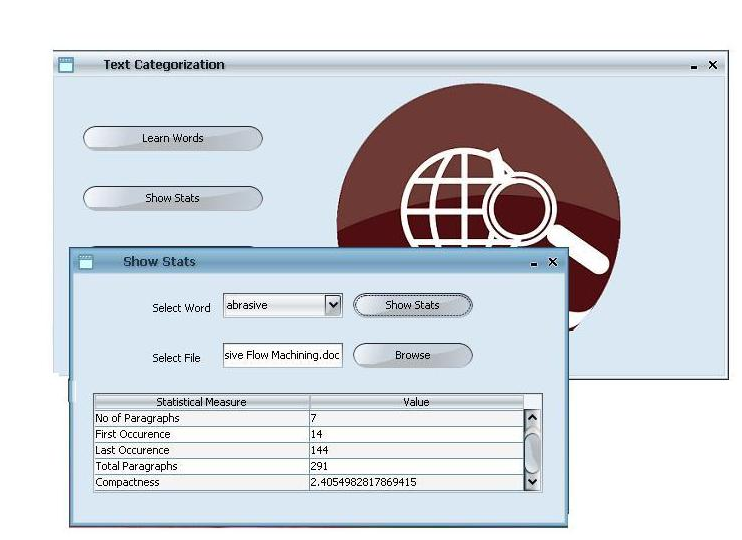}}
\caption{Show stats}
\end{figure}

Figure shows the show stats screen this screen will show the complete results of the analysis done. This screen shows the first appearance, last appearance, compactness and the number paragraphs in the document are displayed.

\begin{figure}
\centering
\resizebox{8cm}{6cm}{\includegraphics{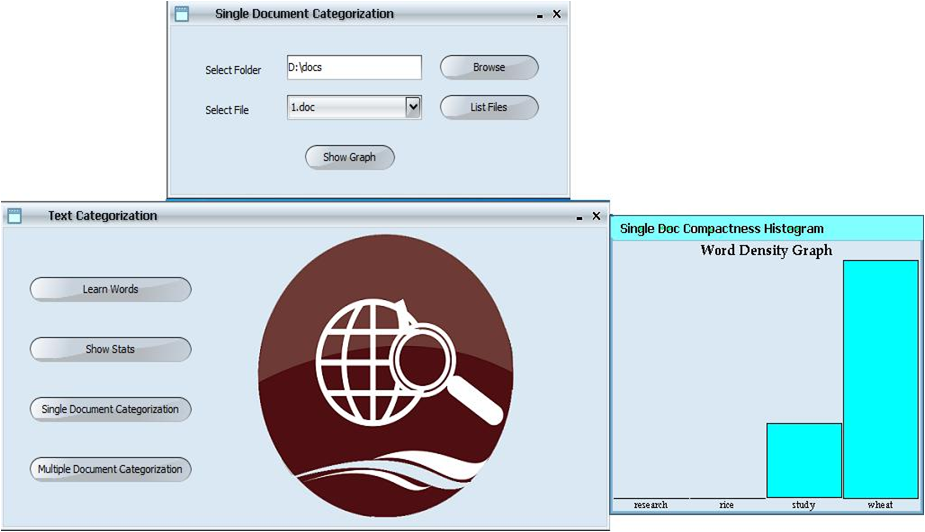}}
\caption{Results of single document compactness histogram}
\end{figure}

Figure shows the results of single document compactness histogram. The important word  from the selected words is displayed.

\begin{figure}
\centering
\resizebox{8cm}{6cm}{\includegraphics{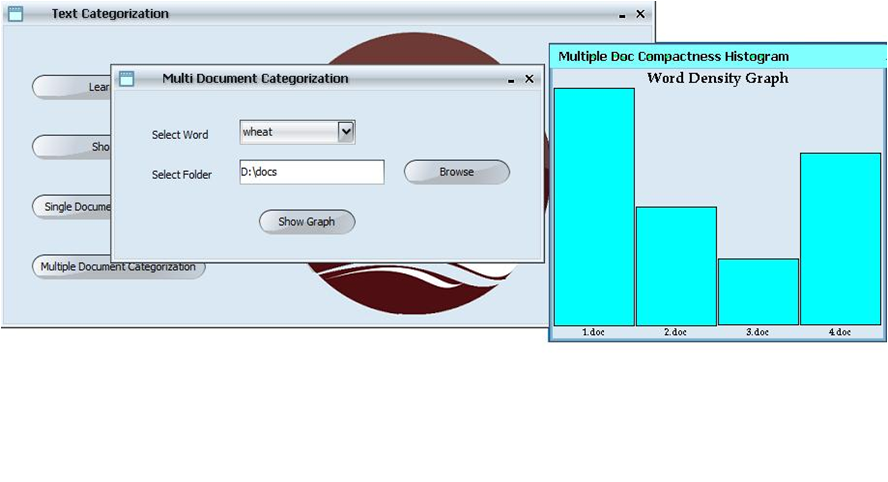}}
\caption{Results of multiple document compactness histogram}
\end{figure}

Figure shows the results of multiple document classification. The document which is more relevant to the selected word is displayed.

\section {Conclusions}
The Classification of the documents should be done by using the distributional features.The measures should be used and all the features of the documents are to be extracted.The extracted features should be observed and the right decision should be taken in classification of document.

The right combination of the features should be taken to gain the full advantages of the Distributional features.The Features along with good classification algorithm should be used for classification of text.

\section{The References Section}\label{references}

\noindent{\includegraphics[width=1in,height=1.7in,clip,keepaspectratio]{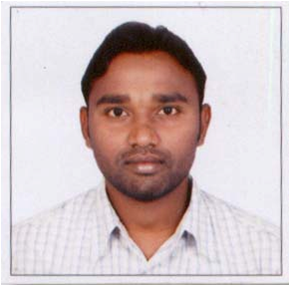}}
\begin{minipage}[b][1in][c]{1.8in}

{\centering{\bf {Srikanth Bethu}} is currently the Assistant Professor, Holy Mary Institute of Technology and Science, JNTU Hyderabad, Hyderabad. He obtained his Bachelor of Engineering from JNTU Hyderabad. He rece-}\\\\
\end{minipage}
ived his Masters degree in Computer Science and Engineering from Osmania University, Hyderabad. \\\\

\begin{minipage}[b][1in][c]{1.8in}
\noindent{\includegraphics[width=1in,height=1.7in,clip,keepaspectratio]{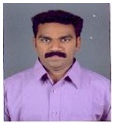}}

{\centering{\bf{G Charless Babu }}is a Professor,Holy Mary Institute of Technology and Science, JNTU Hyderabad, Hyderabad. He was a Professor since 2010 with the Department of Computer Science and Engineering,HITS college,JNTU Hyderabad.}\\\\
\end{minipage}

During the past 10 years of his service at various institutions he has over 30 research publications in refereed International Journals and Conference Proceedings..\\\\

\noindent{\includegraphics[width=1in,height=1.7in,clip,keepaspectratio]{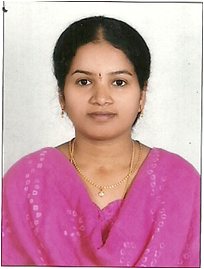}}
\begin{minipage}[b][1in][c]{1.8in}

{\centering{\bf {J Vinoda}} is currently the Assistant Professor, Holy Mary Institute of Technology and Science, JNTU Hyderabad, Hyderabad.}\\\\
\end{minipage}

\noindent{\includegraphics[width=1in,height=1.7in,clip,keepaspectratio]{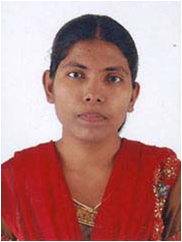}}
\begin{minipage}[b][1in][c]{1.8in}

{\centering{\bf {E Priyadarshini}} is currently the Assistant Professor, Holy Mary Institute of Technology and Science, JNTU Hyderabad, Hyderabad.}\\\\
\end{minipage}

\noindent{\includegraphics[width=1in,height=1.7in,clip,keepaspectratio]{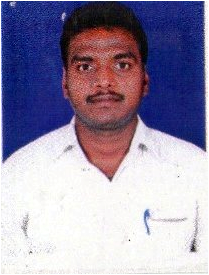}}
\begin{minipage}[b][1in][c]{1.8in}

{\centering{\bf {M Raghavendra rao}} is currently the Assistant Professor, Holy Mary Institute of Technology and Science, JNTU Hyderabad, Hyderabad.}\\\\
\end{minipage}


\begin{thebibliography}{00}

\bibitem{1}
 {Li Youwen,  Xia Shixiong, and Zhou Yong}, "{An Improved KNN Text Classification Algorithm  Based  on  Clustering},"\emph{  Journal  of Computers}, Vol 4, No. 3, pp 230-237, 2009.

\bibitem{2}
 {Chengqing  Zong,  and  Chu-Ren  Huang, Shoushan Li, Rui Xia}, "{A Framework of  Feature Selection Methods for Text Categorization},"\emph{ Proceedings of the 47th Annual  Meeting of the ACL and the 4th IJCNLP of the AFNLP}, pp 	692–700, 2009.

\bibitem{3}
  {F. Li  and  Y. Yang}, "{A Los Function Analysis for Classification Methods  in    Text Categorization},"\emph{  Proc.  20th  Int’l  Conf.  Machine  Learning  (ICML ’03)},  pp  472- 479, 2003.

\bibitem{4}
 {Hyunsoo,  Haesun  Park,  and  Kim  Peg  Howland},  "{Dimension  Reduction  in Text Classification  with   Support   Vector   Machines},"\emph{   Journal  of  Machine  Learning Research}, vol 6, pp 1-17,(2005) .

\bibitem{5}
{Li Baoli, Lu Qin, and Yu Shiwen}, "{An Improved k-Nearest Neighbor Algorithm for Text  Categorization},"\emph{  20th  International  Conference  on  Computer  Processing  of Oriental Languages, Shenyang, China}, pp 1-7, 2003.

\bibitem{6}
 {Li Youwen,  Xia Shixiong, and Zhou Yong} "{An Improved KNN Text Classification Algorithm  Based  on  Clustering},"\emph{  Journal  of Computers}, Vol 4, No. 3, pp 230-237, 2009.

\bibitem{7}
 {Manu  Konchady}, "{Text Mining Application Programming},"\emph{  Cengage Learning },1st edition , pp 209 -233, 2008.

\bibitem{8}
 {N.  Tishby, R.  Bekkerman, R.  El-Yaniv, and Y.  Winter},  "{Distributional   Word  Clusters versus Words for Text Categorization},"\emph{ J. Machine Learning Research}, vol. 3, pp  1182-1208, 2003.

\bibitem{9}
{Xiao-Bing Xue and Zhi-Hua Zhou}, "{Distributional Features for Text Categorization}"  \emph{IEEE Transactons On Knowledge And Data Engineering}, Vol. 21, No. 3, pp 428- 442, 2009.

\bibitem{10}
 {X.-B. Xue and Z.-H. Zhou}, "{Distributional Features for Text Categorization}"  \emph{Proc. 17th European Conf. Machine Learning (ICML ’06)}, pp  497-508, 2006.

 \bibitem{11}
 {X.-B. Xue and Z.-H. Zhou}, "{Distributional Features for Text Categorization}"  \emph{Proc. 17th European Conf. Machine Learning (ICML ’06)}, pp  497-508, 2006.

 \bibitem{12}
 {N.  Tishby, R.  Bekkerman, R.  El-Yaniv, and Y.  Winter},  "{Distributional   Word  Clusters versus Words for Text Categorization},"\emph{ J. Machine Learning Research}, vol. 3, pp  1182-1208, 2003.



\end{thebibliography}
\end{document}